\documentclass[journal]{IEEEtran}
\usepackage{cite}
\usepackage{graphicx}
\usepackage{amsmath}
\usepackage{times}
\usepackage{latexsym}
\usepackage{graphicx}
\usepackage{bm}
\usepackage{amssymb}
\usepackage[center]{caption2}
\usepackage{stfloats}
\usepackage{cases}
\usepackage{array}
\usepackage{setspace}
\usepackage{fancyhdr}
\usepackage{citesort}
\usepackage{color}
\usepackage{subfigure}

\newtheorem{lemma}{Lemma}
\newtheorem{corollary}{Corollary}
\newtheorem{proposition}{Proposition}

\def\proof{\noindent\hspace{2em}{\itshape Proof: }}
\def\endproof{\hspace*{\fill}~$\square$\par\endtrivlist\unskip}

\begin{document}
\title{Wireless Powered Communications: Performance Analysis and Optimization}
\author{Caijun Zhong,~\IEEEmembership{Senior Member,~IEEE,} Xiaoming Chen,~\IEEEmembership{Senior Member,~IEEE,} Zhaoyang Zhang,~\IEEEmembership{Member,~IEEE,} and George K. Karagiannidis,~\IEEEmembership{Fellow,~IEEE}
\thanks{C. Zhong and Z. Zhang are with the Institute of Information and Communication Engineering, Zhejiang University, China. (email: caijunzhong@zju.edu.cn).  Caijun Zhong is also with the National Mobile Communications Research Laboratory, Southeast University,
Nanjing, 210018, China.}
\thanks{X. Chen is with the College of Information and Electronic Engineering, Nanjing University of Aeronautics and Astronautics, Nanjing, China.(email:
chenxiaoming@nuaa.edu.cn).}
\thanks{G. K. Karagiannidis is with Aristotle University of Thessaloniki, Greece, and with Khalifa University, Abu Dhabi, UAE (e-mail: geokarag@ieee.org).}}

\maketitle

\begin{abstract}
This paper investigates the average throughput of a wireless powered communications system, where an energy constrained source, powered by a dedicated power beacon (PB), communicates with a destination. It is assumed that the PB is capable of performing channel estimation, digital beamforming, and spectrum sensing as a communication device. Considering a time splitting approach, the source first harvests energy from the PB equipped with multiple antennas, and then transmits information to the destination. Assuming Nakagami-m fading channels, analytical expressions for the average throughput are derived for two different transmission modes, namely, {\it delay tolerant} and {\it delay intolerant}. In addition, closed-form solutions for the optimal time split, which maximize the average throughput are obtained in some special cases, i.e., high transmit power regime and large number of antennas. Finally, the impact of co-channel interference is studied. Numerical and simulation results have shown that increasing the number of transmit antennas at the PB is an effective tool to improve the average throughput and the interference can be potentially exploited to enhance the average throughput, since it can be utilized as an extra source of energy. Also, the impact of fading severity level of the energy transfer link on the average throughput is not significant, especially if the number of PB antennas is large. Finally, it is observed that the source position has a great impact on the average throughput.

\vspace{1cm}
\begin{center}
{\bf Index Terms}
\end{center}
Energy beamforming, Multiple antennas, Resource allocation, Wireless powered communications.
\end{abstract}

\section{Introduction}
With the rapid evolution of wireless communications systems, handheld mobile devices such as smartphones and tablets, have become one of the primary means to access the Internet. Since these devices are powered by battery with finite capacity, they need to be frequently plugged into the power grid for recharging, which greatly affects the user experience. As such, in an effort to prolong the operational duration of mobile devices, energy harvesting techniques, which scavenge energy from natural resources such as solar and wind, were proposed \cite{V.Raghunathan,B.Medepally}. Ideally, this would allow for perpetual operational time. Nevertheless, the inherent randomness as well as the intermittent property of nature resources makes stable energy output an extremely challenging task. Hence, it may not be suitable for wireless services with stringent quality-of-service (QoS) requirements.

\subsection{Literature}
One of the promising solutions to address the above limitation is to employ radio frequency (RF) signals-based energy harvesting, which can provide stable energy supply, because sufficient level of control over the transmission of RF signals is possible \cite{W.Lumpkins,M.Pinuela}. Since RF signals can carry both energy and information, a new research area, referred to as simultaneous wireless information and power transfer (SWIPT), has recently emerged and has captured the attention of the research community. The pioneering works of Varshney \cite{L.Varshney} and Grover et al. \cite{P.Grover} have investigated the fundamental tradeoff between the capacity and harvested energy for SWIPT systems. Furthermore, practical architectures for SWIPT systems were proposed in \cite{R.Zhang}, where the optimal transmit covariance achieving the rate-energy region was derived. Later in \cite{M.Tao}, the effect of imperfect channel state information (CSI) was considered, while sophisticated architectures improving the rate-energy region were proposed in \cite{L.Liu,X.Zhou}. Moreover, the applications of SWIPT in orthogonal frequency division multiple access (OFDMA) systems and multiuser systems were considered in \cite{Derrick,A.Fou,Trung1,D.S.Mich1}. In addition to the point-to-point communication systems, the application of SWIPT in cooperative relaying networks has also been under extensive investigation \cite{A.Nasir,C.Zhong,Z.Ding,I.Krikidis_2,Z.Ding1,Z.Ding2,I.Krikidis,Z.Ding3,Trung,D.S.Mich,G.Zhu}. Specifically, Nasir et al. in \cite{A.Nasir} studied the throughput performance of an amplify-and-forward (AF) half-duplex relaying system for both time-switching and power-splitting protocols, while \cite{C.Zhong} considered performance of the full-duplex relaying systems. Ding et al. in \cite{Z.Ding} considered the power allocation strategies for decode-and-forward (DF) relaying system with multiple source-destination pairs. A novel and low-complexity antenna switching protocol was proposed in \cite{I.Krikidis_2} to realize SWIPT. Relay selection was considered in \cite{D.S.Mich}, while the impact of multi-antenna relay was investigated in \cite{G.Zhu}, and extension to the two-way relaying was studied in \cite{Trung}. More recently, the performance of energy harvesting cooperative networks with randomly distributed users/relays was studied in \cite{Z.Ding1,Z.Ding2,I.Krikidis}.

In all above-mentioned papers, an integrated base station (BS) architecture is assumed, which acts simultaneously as the information source and energy source. However, due to the huge gap on the operational sensitivity level between the information decoder (in the order of -100 dBm) and the energy harvester (in the order of -10 dBm),  it is only feasible for the BS to power mobile devices within short distance (e.g., less than 10 meters) \cite{K.Huang1}. As such, to enable a full network support for SWIPT, the BS should be deployed in extremely high density (e.g., the cell radius of a BS should be in the range of 10-15 meters), which would incur enormous cost and hence is deemed impractical. Responding to this, Huang and Lau proposed in \cite{K.Huang2} a novel network architecture, where dedicated stations referred to as power beacons (PB) are overlaid with a cellular network to power mobile devices. Since the PB does not require any backhaul connections, the associated cost of PB deployment is much lower, hence, dense deployment of PB to ensure network coverage for SWIPT is practical. In \cite{K.Huang2}, the requirement on the coupled relationship between the PB and BS densities under an outage probability constraint was characterized, which provides useful guidelines for the design of practical wireless powered communications networks.

\subsection{Motivation and Contributions}
Different from the work in \cite{K.Huang2}, which focuses on a large scale network, this work can be viewed as the small scale counterpart. In particular, we consider a point-to-point communication link, where the source is powered by a dedicated PB. The time splitting approach is adopted, i.e., first, the source harvests the energy from the PB, and then performs information transmission with the harvested power. The main objective of this work is to understand the fundamental limit of wireless powered communication links and characterize the impact of key system parameters on the system performance. To this end, we analyze the average throughput of two different transmission modes, namely, {\it delay tolerant} and {\it delay intolerant}. For both modes, we also investigate the optimal time splitting policy maximizing the average throughput. Finally, the impact of co-channel interference is also studied.

The main contributions of the paper can be summarized as follows:
\begin{itemize}
\item Considering the noise limited scenario, we present exact closed-form expressions of the average throughput for both the delay intolerant and the delay tolerant transmission modes. In addition, simplified analytical approximations are also provided for special cases, such as in the high power regime and for large number of antennas, based on which, closed-form solutions are derived for the optimal time split maximizing the average throughput.
\item Considering the interference plus noise scenario, we present an exact integral expression for the average throughput for the delay intolerant transmission mode, and tight throughput bounds for both transmission modes.
\end{itemize}

Numerical and simulation results have shown that putting more antennas at the PB would significantly improve the system throughput. This is rather intuitive, since the more the antennas, the sharper the energy beam, which in turns yields higher energy efficiency. In addition, the findings confirm the dual roles of co-channel interference, i.e., it can be either beneficial or detrimental, depending on the propagation environment and system setup. Moreover, it was revealed that the impact of fading severity level of the energy transfer link on the throughput is not significant, especially if the number of PB antennas is large. Finally, it was shown that the source position has a significant impact on the average throughput. For the noise only scenario, the throughput is a symmetric function of the source position, and the worse case occurs when the source is located at the middle of the PB and destination, while the best case appears when the source moves close to either the PB or destination. On the other hand, for the interference plus noise scenario, the throughput is no longer a symmetric function of the source position, since the optimum position of the source depends on both the source transmit power and interference power.


%
\subsection{Structure and notations}
The remaining of the paper is organized as follows. Section II provides an detailed introduction of the considered system model. Section III presents some preliminary mathematical results. Section IV studies the throughput performance of noise only scenario, while Section V deals with the interference plus noise scenario. Numerical results and discussions are given in Section VI. Finally, Section VII summarizes the paper.

{\it Notation}: We use bold lower case letters to denote vectors and lower case letters to denote scalars. ${\left\| {\bf{h}} \right\|}$ denotes the Frobenius norm; ${\tt E}\{x\}$ stands for the expectation of the random variable $x$; ${*}$ denotes the conjugate operator, while ${\dag}$ denotes the conjugate transpose operator; $\Gamma(x)$ is the gamma function; $K_v(x)$ is the $v$-th order modified Bessel function of the second kind \cite[Eq. (8.407.1)]{Table}; ${\tt Ei}(x)$ is the exponential integral function \cite[Eq. (8.211.1)]{Table}; $\psi \left( x \right)$ is the Digamma function \cite[Eq. (8.360.1)]{Table}; ${\mathop{\rm G}\nolimits}_{p,q}^{m,n} \left( \cdot\right)$ is the Meijer G-function \cite[Eq. (9.301)]{Table}.

\section{System Model}
\begin{figure}[htb!]
\centering
\includegraphics[scale=0.4]{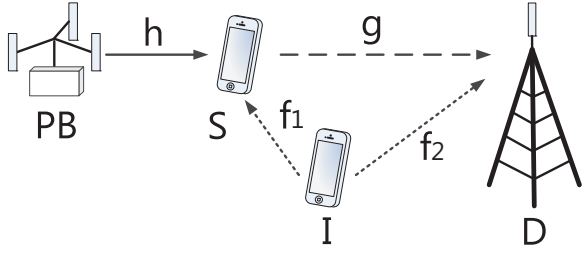}
\caption{System model: PB, S, D, and I denote the power beacon, source, destination, and interferer, respectively.}\label{fig:fig0}
\end{figure}

We consider a point-to-point communication system where the source S communicates with the destination D, as depicted in Fig. \ref{fig:fig0}. We assume that S is an energy constrained mobile node, hence relies on the external energy charging via wireless power transfer from a dedicated PB.\footnote{In the current work, PB only supplies wireless energy to the source, and does not participate in the information transmission. However, similar to \cite{Y.Zeng00,Y.Zeng01}, it is assumed that the PB is empowered with the functionality of a communication entity, and is capable of performing tasks such as channel estimation, digital beamforming, and spectrum sensing, etc.} Both S and D are equipped with a single antenna, while the PB is equipped with $N$ antennas. Full channel state information of the link between the PB and S is assumed at the PB. In practice, the channel can be estimated by overhearing the pilot sent by the source.

Assuming a block time of $T$, during the first phase of duration $\tau T$, where $0<\tau<1$, S harvests energy from the PB. In the remaining time of duration $(1-\tau)T$, S transmits information to D. Please note, such a two-stage communication protocol has also been considered in prior works such as \cite{H.Chen}, where it was termed as the ``harvest-then-transmit'' protocol. We now consider two separate cases depending on the existence of co-channel interference.

\subsection{Noise Limited Scenario}
For the noise limited case, during the energy harvesting phase, the received signal at S can be expressed as \cite{R.Zhang,G.Yang}
\begin{align}
y_s = \sqrt{\frac{P}{d_1^\alpha}}{\bf h}{\bf x}_s+n_s,
\end{align}
where $P$ is the transmit power at the PB, $d_1$ denotes the distance between PB and S, $\alpha$ is the path loss exponent, ${\bf x}_s$ is an $N\times 1$ signal vector, and $n_s$ is the additive white Gaussian noise (AWGN) with ${\tt E}\{n_sn_s^*\} = N_0$.

Due to the relatively short power transfer distance, the line-of-sight path is likely to exist between the PB and S, hence, it is a natural choice to use the Rician distribution in order to model the PB to S channel. However, the complicated Rician fading probability density function (pdf) makes the ensuing analysis extremely difficult. On the other hand, it is well known that the Nakagami-m fading distribution provides a very good approximation to the Rician distribution. Motivated by this, and to simplify the analysis, we adopt the Nakagami-m fading to model the PB to S channel in this paper. Therefore, the elements of $\mathbf{h}=[h_{i}]$, $i = 1, \cdots, N$, are assumed to be independent and identically distributed (i.i.d.)
with uniformly distributed phase and the magnitude, $x=|h_{i}|$, following a Nakagami-$m$ pdf
\begin{equation}\label{Eqn:Nakagami}
p(x)=\frac{2}{\Gamma(m)}\left(\frac{m}{\Omega}\right)^{m}x^{2m-1}e^{-\left(\frac{m}{\Omega}\right)x^2},~~\mbox{$x\geq 0$},
\end{equation}
where $\Gamma(\cdot)$ denotes the gamma function, $m\triangleq\frac{{\tt E}^2[x^2]}{{\tt Var}[x^2]}$, and $\Omega\triangleq{\tt
E}[x^2]$. Without loss of generality, we set $\Omega=1$.

Since the PB is equipped with multiple antennas, energy beamforming is applied to improve the efficiency of energy transfer, i.e.,
\begin{align}
{\bf x}_s={\bf w}s_e,
\end{align}
where ${\bf w}$ is the beamforming vector with $\|{\bf w}\|^2 =1$, while $s_e$ is the energy symbol with unit power. It is easy to observe that the optimal beamforming vector is given by
\begin{align}
{\bf w} = \frac{{\bf h}^{\dag}}{\|{\bf h}\|}.
\end{align}
As such, the total received energy at the end of the first phase can be computed as
\begin{align}
E_n = \frac{\eta\|{\bf h}\|^2P\tau T}{d_1^{\alpha}},
\end{align}
where $0<\eta<1$ is the energy conversion efficiency.

In the second phase, S transmits information to D using the energy harvested in the first phase. Hence, the received signal $y_D$ at D is given by
\begin{align}
y_D=\sqrt{\frac{E_n}{(1-\tau)Td_2^{\alpha}}}gs_0+n_d,
\end{align}
where $d_2$ denotes the distance between S and D, $g$ is the channel coefficient following complex Gaussian distribution with zero-mean and unit variance, $s_0$ is the information symbol with unit energy, and $n_d$ is the AWGN with variance $N_0$. Therefore, the end-to-end signal to noise ratio (SNR) can be computed as
\begin{align}\label{snr:noise}
\gamma_N = \frac{\tau\eta\|{\bf h}\|^2|g|^2P}{(1-\tau)d_1^{\alpha}d_2^{\alpha}N_0}.
\end{align}

\subsection{Interference plus Noise Scenario}
In the current work, we consider the scenario with a single dominant co-channel interferer. It is worth pointing out that the single dominant interferer assumption has been widely adopt in the literature, see \cite{R.Mallik,M.Hassanien} and references therein. Moreover, such a system model enables us to gain key insights on the joint effect of path loss exponent, network topology and interference power in a wireless powered network.\footnote{Although the single dominant interferer assumption is adopted, noticing that the sum of independent and non-identically distributed exponential random variables follows hyper-exponential distribution, the ensuing analysis could actually be extended to the more general multiple interferers scenario by following almost the same derivations.} In the presence of co-channel interference, the received signal during the energy harvesting phase is given by
\begin{align}
y_s = \sqrt{\frac{P}{d_1^\alpha}}{\bf h}{\bf x}_s+\sqrt{\frac{P_I}{d_3^{\alpha}}}f_1s_i+n_s,
\end{align}
where $P_I$ is the transmit power of the interferer,\footnote{Please note, we assume that the interference power remains constant during the entire communication block $T$.} $d_3$ denotes the distance between I and S, $f_1$ is the channel coefficient following complex Gaussian distribution with zero-mean and unit variance, $s_i$ is the interference symbol with unit energy. As such, the total received energy at the end of the first phase can be computed as \cite{L.Liu}
\begin{align}
E_I = \frac{\eta\|{\bf h}\|^2P\tau T}{d_1^{\alpha}}+\frac{\eta|f_1|^2P_I\tau T}{d_3^{\alpha}}.
\end{align}

In the second phase, S transmits the signal to D, and the received signal $y_D$ at D is given by
\begin{align}
y_D=\sqrt{\frac{E_I}{(1-\tau)Td_2^{\alpha}}}gs_0+\sqrt{\frac{P_I}{d_4^{\alpha}}}f_2s_i+n_d,
\end{align}
where $d_4$ denotes the distance between I and D, and $f_2$ is the channel coefficient following complex Gaussian distribution with zero-mean and unit variance. Therefore, the end-to-end signal to interference plus noise ratio (SINR) can be computed as
\begin{align}
\gamma_I =\frac{ \frac{\tau\eta|g|^2}{(1-\tau)d_2^{\alpha}}\left(\frac{\|{\bf h}\|^2P}{N_0d_1^{\alpha}}+
\frac{|f_1|^2P_I}{N_0d_3^{\alpha}}\right)}{1+\frac{P_I|f_2|^2}{N_0d_4^{\alpha}}}.
\end{align}


\section{Preliminaries}
In this section, we present some preliminary results on the solutions of some optimization problems, which will be frequently invoked in the analysis.
\begin{lemma}\label{lemma:1}
Consider the following function
\begin{align}
g(x) = (1-x)\left(1-b\frac{1-x}{x}\right),\quad 0<x<1,
\end{align}
where $b$ is a positive real number, then the optimal $x^*$ which maximizes $g(x)$ is given by
\begin{align}
x^* = \sqrt{\frac{b}{b+1}}.
\end{align}
\end{lemma}
\proof
It is easy to show that the second derivative of $g(x)$ with respect to $x$ is given by
\begin{align}
g''(x) = -\frac{2b}{x^3},
\end{align}
which is strictly smaller than zero. Hence, $g(x)$ is a concave function with respect to $x$. Therefore, a unique $x^*$ which maximizes $g(x)$ exists, and can be computed by solving
\begin{align}
g'(x) = -1-b+\frac{b}{x^2} = 0.
\end{align}
This completes the proof.
\endproof

\begin{lemma}\label{lemma:2}
Consider the following function
\begin{align}
g(x) = (1-x)\exp\left(-\frac{b}{x}\right),\quad 0<x<1,
\end{align}
where $b$ is a positive real number, then the optimal $x^*$ which maximizes $g(x)$ is given by
\begin{align}
x^* = \frac{\sqrt{b^2+4b}-b}{2}.
\end{align}
\end{lemma}
\proof See Appendix \ref{app:lemma:2}.\endproof

\begin{lemma}\label{lemma:3}
Consider the following function
\begin{align}
g(x) = (1-x)\ln\left(1+\frac{bx}{1-x}\right),\quad 0<x<1,
\end{align}
where $b$ is a positive real number, then the optimal $x^*$ which maximizes $g(x)$ is given by
\begin{align}
x^* = \frac{e^{W\left(\frac{b-1}{e}\right)+1}-1}{b+e^{W\left(\frac{b-1}{e}\right)+1}-1},
\end{align}
where $W(x)$ is the Lambert W function \cite{Wfun}.
\end{lemma}
\proof See Appendix \ref{app:lemma:3}.
\endproof

\begin{lemma}\label{lemma:4}
Consider the following function
\begin{align}
g(x) = (1-x)\left(\ln\frac{x}{1-x}+a\right), \quad 0<x<1,
\end{align}
where $a$ is a real number, then the optimal $x^*$ which maximizes $g(x)$ is given by
\begin{align}
x^* = \frac{e^{W\left(e^{a-1}\right)+1-a}}{e^{W\left(e^{a-1}\right)+1-a}+1}.
\end{align}
\end{lemma}
\proof
It can be shown that the second derivative of $g(x)$ with respect to $x$ is given by
\begin{align}
g''(x) = -\frac{1}{x^2}-\frac{1}{x(1-x)},
\end{align}
which is strictly smaller than zero, hence $g(x)$ is a concave function with respect to $x$. Therefore, the optimal $x^*$ which maximizes $g(x)$ can be obtained by
\begin{align}
g'(x) = -a +\frac{1}{x}-\ln\frac{x}{1-x} = 0
\end{align}
The desired result can then be obtained by following the similar manipulations as in the proof of Lemma \ref{lemma:3}.
\endproof

\section{Noise Limited Scenario}
In this section, we focus on the noise limited scenario, and study the achievable average throughput of wireless powered communication network.

\subsection{Delay intolerant transmission}
For delay intolerant transmission, the source transmits at a constant rate $R_c$, which may be subjected to outage due to fading. Hence, the average throughput can be evaluated as
\begin{equation}\label{throughput:dc}
R_{\sf DC}(\tau)=(1-P_{\sf out})R_c(1-\tau),
\end{equation}
where $P_{\sf out}$ is the outage probability.

\begin{proposition}\label{prop:1}
When the source transmission rate is $R_c$, then the average throughput of the system is given by
\begin{multline}
R_{\sf DC}(\tau) = \frac{2R_cm^{\frac{Nm}{2}}}{\Gamma(Nm)}(1-\tau)\left(\frac{(1-\tau)c_1 \gamma_{\sf th}}{\tau}\right)^{\frac{Nm}{2}}\\K_{Nm}\left(2\sqrt{\frac{(1-\tau)mc_1 \gamma_{\sf th}}{\tau}}\right),
\end{multline}
where $\Gamma(x)$ is the Gamma function \cite[Eq. (8.310)]{Table}, $K_n(x)$ is the modified Bessel function of the second kind \cite[Eq. (8.432)]{Table}, $c_1 = \frac{d_1^{\alpha}d_2^{\alpha}N_0}{\eta P}$ and $\gamma_{\sf th} = 2^{R_c}-1$.
\end{proposition}
\proof The key is to obtain the outage probability of the system, which can be written as
\begin{align}
P_{\sf out} &= {\sf Pr}\left\{\gamma_N<\gamma_{\sf th}\right\}={\sf Pr}\left\{\|{\bf h}\|^2|g|^2<\frac{(1-\tau)c_1 \gamma_{\sf th}}{\tau}\right\}.
\end{align}
According to \cite{M.Evans}, $\|{\bf h}\|^2$ is a Gamma random variable with pdf given by
\begin{align}
p(x) = \frac{m^{Nm}}{\Gamma(Nm)}x^{Nm-1}e^{-mx}, \mbox{ for } x\geq 0.
\end{align}
Please note that the cumulative distribution function (cdf) of $\|{\bf h}\|^2|g|^2$ can be obtained by invoking \cite[Eq. 8]{S.Jin} with some algebraic manipulations. However, the resulting expression involves multiple summations. Here we adopt a slightly different approach to obtained a simple alternative cdf expression. Conditioned on $\|{\bf h}\|^2$, the cdf of $\|{\bf h}\|^2|g|^2$ is given by
\begin{align}
F(x|\|{\bf h}\|^2) =1-e^{-\frac{x}{\|{\bf h}\|^2}}.
\end{align}
To this end, averaging over $\|{\bf h}\|^2$, with the help of \cite[Eq. 3.471.9]{Table}, the unconditional cdf can be computed as
\begin{align}\label{snrcdf}
F(x) = 1-\frac{2(mx)^{\frac{Nm}{2}}}{\Gamma(Nm)}K_{Nm}\left(2\sqrt{xm}\right).
\end{align}
The desired result can be then obtained after some simple algebraic manipulations.
\endproof
%
Having characterized the average throughput of the system, the optimal time split $\tau$ could be obtained by solving the following optimization problem
\begin{align}
&\tau^* = \mbox{arg} \max_{\tau}R_{\sf DC}(\tau)\notag\\
&\mbox{s.t. }0<\tau<1.
\end{align}
In general, due to the complexity of the involved expression, obtaining an exact closed-form solution of $\tau^*$ is very challenging. However, it can be efficiently solved by a one-dimensional search, or instead, it can be numerically evaluated using the build-in function ``NSlove'' of Mathematica.

Next, to gain more insights, we now look into some special cases, where closed-form solutions of $\tau^*$ can be obtained.
\begin{corollary}\label{coro:1}
In the high transmit power regime, i.e., $P\rightarrow \infty$, $\tau^*$ is given by
\begin{align}\label{eqn:largesnr}
\tau^* =\sqrt{\frac{mc_1\gamma_{\sf th}}{mc_1\gamma_{\sf th}+Nm-1}},
\end{align}
where $c_1$ and $\gamma_{\sf th}$ is given in Proposition \ref{prop:1}.
\end{corollary}
\proof
When $P$ is sufficiently large, using the asymptotic expansion of $K_n(x)$ given in \cite[Eq. 9.6.11]{Handbook}, the outage probability can be accurately approximated by
\begin{align}\label{eqn:highout}
P_{\sf out} \approx \frac{m}{mN-1}\frac{(1-\tau)c_1 \gamma_{\sf th}}{\tau}.
\end{align}
Substituting (\ref{eqn:highout}) into (\ref{throughput:dc}), we have
\begin{align}\label{eqn:rdc}
R_{\sf DC}(\tau) =2 R_c(1-\tau)\left(1-\frac{m}{mN-1}\frac{(1-\tau)c_1 \gamma_{\sf th}}{\tau}\right).
\end{align}
To this end, invoking Lemma \ref{lemma:1} yields the desired result.
\endproof

Eq. (\ref{eqn:rdc}) is a simple expression which can be efficiently used to analyze the impact of critical system parameters, such as transmit power $P$, number of antennas $N$ and transmission rate $R_c$ on the optimal time split. \\
{\bf Remark 1}: $\tau^*$ is a monotonically decreasing function of $P$ and $N$. This can be easily concluded when the transmit power $P$ is large, to harvest the same amount of power, much less time is needed. Similarly, when the number of antennas $N$ is large, the energy beamforming gain increases sharply, which also reduces the required time for energy harvesting.\\
{\bf Remark 2}: $\tau^*$ is a monotonically decreasing function of the Nakagami-m fading parameter $m$. Since a larger $m$ implies less severer fading environment, this indicates that shorter energy harvesting time is needed if the fading condition improves. In addition, when $m\rightarrow \infty$, i.e., the channel becomes deterministic, the optimal $\tau^*=\sqrt{\frac{c_1\gamma_{\sf th}}{c_1\gamma_{\sf th}+N}}$.\\
{\bf Remark 3}: As expected, $\tau^*$ is a monotonically increasing function of $R_c$. When $R_c$ becomes large, to enable reliable communication at a higher data rate, a larger transmit power is required, hence, more time is needed for energy harvesting.

Next we consider the scenario where the number of antennas is large.
\begin{corollary}
In the asymptotically large number of antennas regime, i.e., $N\rightarrow \infty$, the optimal $\tau^*$ is given by
\begin{align}\label{eqn:largeN}
\tau^*=\frac{1}{2}\left({\sqrt{\left(\frac{c_1\gamma_{\sf th}}{N}\right)^2+\frac{4c_1\gamma_{\sf th}}{N}}-\frac{c_1\gamma_{\sf th}}{N}}\right),
\end{align}
where $c_1$ and $\gamma_{\sf th}$ is given in Proposition \ref{prop:1}.
\end{corollary}
\proof
In the asymptotically large number of antennas regime, according to the law of large numbers, we have
\begin{align}
\|{\bf h}\|^2 \approx  N.
\end{align}
As such, the outage probability of the system can be approximated as
\begin{align}\label{eqn:largen}
P_{\sf out} &\approx {\sf Pr}\left(|g|^2<\frac{(1-\tau)c_1 \gamma_{\sf th}}{N\tau}\right)\notag\\
&=1-\exp\left(-\frac{(1-\tau)c_1 \gamma_{\sf th}}{N\tau}\right).
\end{align}
Substituting (\ref{eqn:largen}) into (\ref{throughput:dc}), we have
\begin{align}
R_{\sf DC}(\tau) =R_c\exp\left(\frac{c_1 \gamma_{\sf th}}{N}\right)(1-\tau)\exp\left(-\frac{c_1 \gamma_{\sf th}}{N\tau}\right).
\end{align}
To this end, invoking Lemma \ref{lemma:2} yields the desired result.
\endproof

Now, if $f(x,y)=\sqrt{\left(\frac{x}{y}\right)^2+\frac{4x}{y}}-\frac{x}{y}$, then
 \begin{align}
 &\frac{df(x,y)}{dx} = \frac{1}{y}\left(\left(\frac{x}{y}+2\right)\left(\left(\frac{x}{y}\right)^2+\frac{4x}{y}\right)^{-1/2}-1\right)\\
 &> \frac{1}{y}\left(\left(\frac{x}{y}+2\right)\left(\left(\frac{x}{y}\right)^2+\frac{4x}{y}+4\right)^{-1/2}-1\right)=0.
 \end{align}
Similarly, taking the first derivative of $f(x,y)$ with respect to $y$, it can be easily shown that $\frac{df(x,y)}{dy}<0$. As such, once again we can establish that $\tau^*$ is a monotonically decreasing function of $P$ and $N$, and is an monotonically increasing function of $R_c$.

%
%
%

\subsection{Delay tolerant transmission}
In the delay tolerant transmission scenario, the source transmits at any constant rate upper bounded by the ergodic capacity. Since the codeword length is sufficiently large compared to the block time, the codeword could experience all possible realizations of the channel. As such, the ergodic capacity becomes an appropriate measure. Hence, the throughput of the system is given by
\begin{align}
R_{\sf DT} = (1-\tau)C_{e},
\end{align}
where $C_{e}$ is the ergodic capacity of the system.

\begin{proposition}\label{prop:2}
The average throughput of the system is given by
\begin{align}\label{eqn:dtt}
R_{\sf DT}(\tau) = \frac{(1-\tau)\log_2e}{\Gamma(mN)} G_{4,2}^{1,4}\left(\left.\frac{\tau }{(1-\tau)c_1m}\right|_{1,0}^{1-mN,0,1,1}\right),
\end{align}
where $G_{p,q}^{m,n}(x)$ is the Meijer G-function \cite[Eq. 9.301]{Table}.
\end{proposition}
\proof
Starting from the definition, the ergodic capacity can be computed as
\begin{align}
C_e &= {\tt E}\left\{\log_2\left(1+\gamma_N\right)\right\}.
\end{align}
Now, from the cdf expression given in Eq. (\ref{snrcdf}), with the help of the derivative property of bessel K function \cite[Eq. 9.6.28]{Handbook}, the pdf of $\|{\bf h}\|^2|g|^2$ can be obtained as
\begin{align}
f(x) = \frac{2m^{Nm}}{\Gamma(Nm)}\left(\frac{x}{m}\right)^{\frac{mN-1}{2}}K_{mN-1}(2\sqrt{mx}).
\end{align}
Hence, we have
\begin{align}
&C_e=\frac{2\log_2e }{\Gamma(Nm)}\times\notag\\
&\int_0^{\infty}\ln\left(1+\frac{\tau x}{(1-\tau)mc_1}\right)x^{\frac{mN-1}{2}}K_{mN-1}(2\sqrt{x})dx.
\end{align}
To proceed, we first express the logarithm in terms of a Meijer G-function as \cite[Eq. 8.4.6.5]{A.Prud}
\begin{align}
\ln\left(1+\frac{\tau x}{(1-\tau)mc_1}\right) = G_{2,2}^{1,2}\left(\left.\frac{\tau x}{(1-\tau)mc_1}\right|_{1,0}^{1,1}\right),
\end{align}
then the desired result can be obtained after some algebraic manipulations with the help \cite[Eq. 7.821.3]{Table}.
\endproof

Having obtained the average throughput of the system, the optimal time split $\tau$ could be found by solving the following optimization
\begin{align}
&\tau^* = \mbox{arg} \max_{\tau}R_{\sf DT}(\tau)\notag\\
&\mbox{s.t. }0<\tau<1.
\end{align}
Due to the presence of Meijer G-function, it is hard to derive an exact closed-form solution for $\tau^*$. Motivated by this, in the following, we present a simple lower bound for the throughput $R_{\sf DT}^l(\tau)$.
\begin{corollary}
The average throughput of the system is lower bounded by
\begin{multline}\label{eqn:lower1}
R_{\sf DT}^l(\tau)=\\(1-\tau) \log_2\left(1+\frac{\tau \exp\left\{\psi(mN)-\ln m+\psi(1)\right\}}{(1-\tau)c_1}\right),
\end{multline}
and the optimal $\tau^*$ which maximizes the lower bound $R_{\sf DT}^l$ is given by
\begin{align}
\tau^*=\frac{e^{W\left(\frac{a-1}{e}\right)+1}-1}{a+e^{W\left(\frac{a-1}{e}\right)+1}-1},
\end{align}
where $a = \frac{\exp\left\{\psi(mN)-\ln m+\psi(1)\right\}}{c_1}$, $\psi(x)$ is the digamma function \cite[Eq. 8.360]{Table} and $W(x)$ is the Lambert W function.
\end{corollary}
\proof Using the fact that $\log(1+\exp(x))$ is a convex function with regard to $x$, the ergodic capacity can be lower bounded by
\begin{align}
C_e &\geq \log_2\left(1+\exp\left\{{\tt E}\left\{\ln\gamma\right\}\right\}\right)\notag\\
&=\log_2\left(1+\frac{\tau }{(1-\tau)c_1}\exp\left\{{\tt E}\left\{\ln\|{\bf h}\|^2+\ln|g|^2\right\}\right\}\right).\notag
\end{align}
Utilizing \cite[Eq. 4.352]{Table},
we obtain
\begin{align}
{\tt E}\left\{\ln\|{\bf h}\|^2\right\}& = \frac{m^{Nm}}{\Gamma(Nm)}\int_0^{\infty}\ln x x^{mN-1}e^{-mx}dx \notag\\
&=\psi(mN)-\ln m,
\end{align}
which completes the first half of the proof. As for $\tau^*$, it can be obtained by invoking Lemma \ref{lemma:3}.
\endproof

Similarly, we now look at the high SNR regime, and we have
\begin{corollary}
In the high SNR regime, i.e., $P\rightarrow \infty$,  the average throughput can be approximated by
\begin{align}
R_{\sf DC}&\approx \frac{1-\tau}{\ln 2}\times\notag\\
&\left(\ln\frac{\tau}{1-\tau}+\ln\frac{\eta P}{d_1^{\alpha}d_2^{\alpha}N_0}+\psi(mN)-\ln m+\psi(1)\right),\notag
\end{align}
and the corresponding $\tau^*$ is given by
\begin{align}
\tau^*= \frac{e^{W\left(e^{a-1}\right)+1-a}}{e^{W\left(e^{a-1}\right)+1-a}+1},
\end{align}
with $a$ being
\begin{align}
a=\ln\frac{\eta P}{d_1^{\alpha}d_2^{\alpha}N_0}+\psi(mN)-\ln m+\psi(1).
\end{align}
\end{corollary}
\proof
In the high SNR regime, the ergodic capacity can be approximated by
\begin{align}
&C_e \approx {\tt E}\left\{\log_2 \gamma_N\right\}\notag\\
&=\frac{1}{\ln 2}\left(\ln\frac{\tau}{1-\tau}+\ln\frac{\eta P}{d_1^{\alpha}d_2^{\alpha}N_0}+{\tt E}\left\{\ln\|{\bf h}\|^2\right\}+{\tt E}\left\{\ln|g|^2\right\}\right)\notag\\
&=\frac{1}{\ln 2}\left(\ln\frac{\tau}{1-\tau}+\ln\frac{\eta P}{d_1^{\alpha}d_2^{\alpha}N_0}+\psi(mN)-\ln m+\psi(1)\right).\notag
\end{align}
To this end, invoking Lemma \ref{lemma:4} yields the desired result.
\endproof

\section{Interference plus Noise Scenario}
In this section, we consider the interference plus noise scenario. For both transmission modes, analytical expressions for the average throughput are presented.

\subsection{Delay intolerant transmission}

\begin{proposition}\label{prop:3}
When the source transmission rate is $R_c$, then the average throughput of the system is given by (\ref{eq:rdc}) shown on the top of the next page,
\begin{figure*}
\begin{multline}\label{eq:rdc}
R_{\sf DC}(\tau) =\frac{R_c(1-\tau)m^{Nm}}{\rho_1^{Nm}\rho_I}\left(\sum_{t=1}^{Nm}\frac{(-1)^{t-1}}{(Nm-t)!}\left(\frac{1}{\rho_I}-\frac{m}{\rho_1}\right)^{-t}\int_0^{\infty}\frac{x^{Nm-t+1}}{x+\frac{\gamma_{\sf th}}{b_2}}e^{-\frac{mx}{\rho_1}-\frac{b_1\gamma_{\sf th}}{b_2 x}}dx\right.\\
\left.+\left(\frac{m}{\rho_1}-\frac{1}{\rho_I}\right)^{-Nm}\int_0^{\infty}\frac{xe^{-\frac{x}{\rho_I}-\frac{b_1\gamma_{\sf th}}{b_2 x}}}{x+\frac{\gamma_{\sf th}}{b_2}}dx\right).
\end{multline}
\hrule
\end{figure*}
where $\rho_1=\frac{P}{N_0d_1^{\alpha}}$, $\rho_I=\frac{P_I}{N_0d_3^{\alpha}}$, $b_1 = \frac{N_0d_4^{\alpha}}{P_I}$, and $b_2 = \frac{\tau\eta N_0 d_4^{\alpha}}{(1-\tau)d_2^{\alpha}P_I}$.
\end{proposition}
\proof See Appendix \ref{app:prop:3}.\endproof

%

Proposition \ref{prop:3} provides an exact integral expression for the average throughput of the system, which may not be amenable for further manipulations. Alternatively, the following upper bound on the average throughput can be efficiently used.
\begin{corollary}\label{coro:5}
When the source transmission rate is $R_c$, the average throughput of the system is upper bounded by
\begin{align}\label{eqn:coro:5}
R_{\sf DC}(\tau)\leq R_{\sf DC}^u(\tau)=R_c(1-\tau)(1-P_{\sf out}^l),
\end{align}
where $P_{\sf out}^l$ is given by (\ref{eqn:poutl}) shown on the top of the next page,
\begin{figure*}
\begin{multline}\label{eqn:poutl}
P_{\sf out}^l=\frac{\gamma_{\sf th}m^{Nm}}{b_2\rho_1^{Nm}\rho_I}\left(\sum_{t=1}^{Nm}\frac{(-1)^{t-1}}{(Nm-t)!}\left(\frac{1}{\rho_I}-\frac{m}{\rho_1}\right)^{-t}\left((-1)^{Nm-t-1}\left(\frac{\gamma_{\sf th}}{b_2}\right)^{Nm-t}e^{\frac{\gamma_{\sf th}m}{b_2\rho_1}}{\tt Ei}\left(-\frac{\gamma_{\sf th}m}{b_2\rho_1}\right)\right.\right.\\
\left.\left.+\sum_{k=1}^{Nm-t}(k-1)!\left(-\frac{\gamma_{\sf th}}{b_2}\right)^{Nm-t-k}\left(\frac{\rho_1}{m}\right)^k\right)-\left(\frac{m}{\rho_1}-\frac{1}{\rho_I}\right)^{-Nm}e^{\frac{\gamma_{\sf th}}{b_2\rho_I}}{\tt Ei}\left(-\frac{\gamma_{\sf th}}{b_2\rho_I}\right)\right).
\end{multline}
\hrule
\end{figure*}
and ${\tt Ei}(x)$ is the exponential integral function \cite[Eq. 8.211]{Table}.
\end{corollary}
\proof
The key is to notice that the exact end-to-end SINR $\gamma_I$ can be tightly bounded by $\gamma_I^u$ as
\begin{align}
\gamma_I\leq \gamma_I^u =b_2ZV,
\end{align}
where $Z = \frac{|g|^2}{|f_2|^2}$.
Please note, $\gamma_I^u$ could also be interpreted as the signal to interference ratio in an interference-limited scenario. It is easy to show that the cdf of the random variable (RV) $Z$ is given by
\begin{align}
F_Z(x) = \frac{x}{1+x}.
\end{align}
Hence, utilizing the pdf of $V$ given in (\ref{pdf:v}), 
and with the help of the integration formula \cite[Eq. 3.354.5]{Table}, we obtain the desired result.
\endproof



Having characterized the average throughput of the system, the optimal time split $\tau^*$ can be obtained by solving the following optimization problem
\begin{align}
&\tau^* = \mbox{arg} \max_{\tau}R_{\sf DC}^u(\tau)\notag\\
&\mbox{s.t. }0<\tau<1.
\end{align}
In general, due to the complexity of the involved expression, obtaining an exact closed-form solution of $\tau^*$ is very challenging. However, a one-dimensional search can used to efficiently obtain the optimal solution. Or one can simply invoke the build-in function ``NSlove'' of Mathematica to get $\tau^*$.

\subsection{Delay tolerant transmission}
When the delay tolerant transmission mode is considered, we present the following key result.
\begin{proposition}\label{prop:6}
If the delay tolerant transmission mode is considered, the average throughput of the system can be lower bounded as
\begin{align}\label{eqn:lower2}
R_{DL}(\tau)\geq R_{DL}^l(\tau)=(1-\tau)\log_2\left(1+ \frac{a\tau}{1-\tau}\right)
\end{align}
and the optimal $\tau^*$ maximizing the lower bound $R_{DL}^l(\tau)$ can be computed via
\begin{align}
\tau^*=\frac{e^{W\left(\frac{a-1}{e}\right)+1}-1}{a+e^{W\left(\frac{a-1}{e}\right)+1}-1},
\end{align}
where $a$ is given by
\begin{align}
a=&\exp\left\{\psi(1)+e^{b_1}{\tt Ei}(-b_1)+\ln\frac{\eta}{d_2^{\alpha}}+a_0\right\},
\end{align}
with
\begin{figure*}
\begin{multline}
a_0=\frac{m^{Nm}}{\rho_1^{Nm}\rho_I}\left(\sum_{t=1}^{Nm}\frac{(-1)^{t-1}}{(Nm-t)!}\left(\frac{1}{\rho_I}-\frac{m}{\rho_1}\right)^{-t}\left(\left(\frac{\rho_1}{m}\right)^{mN-t+1}\Gamma(mN-t+1)\left(\psi(mN-t+1)+\ln\frac{\rho_1}{m}\right)\right)\right.\\
\left.+\left(\frac{m}{\rho_1}-\frac{1}{\rho_I}\right)^{-Nm}\left(\rho_I\left(\psi(1)+\ln\rho_I\right)\right)\right).
\end{multline}
\hrule
\end{figure*}
\end{proposition}
\proof See Appendix \ref{app:prop:6}.\endproof

\section{Numerical and Simulation Results}
In this section, Monte Carlo simulation results are presented to validate the analytical expressions derived in the previous sections. All the simulation results are obtained by averaging over $10^6$ independent trials. Unless otherwise specified, the following set of parameters were used in simulations: $R_c = 1 \mbox{ bps/Hz}$, hence the outage SNR threshold is given by $\gamma_{\sf th} = 2^{R_c}-1=1$. The energy conversion efficiency is set to be $\eta = 0.4$, while the path loss exponent is set to be $\alpha=2.5$, the Nakagami-m parameter is set to be $m=4$, which corresponds to a Rician factor of $K=3+\sqrt{12}$. The distances between the PB and S, S and D are set to be $d_1 = 8$m and $d_2 = 15$m, respectively.

\begin{figure}[ht]
  \centering
  \subfigure[Noise only: $\tau=0.5$.]{\label{fig:2a}\includegraphics[width=0.45\textwidth]{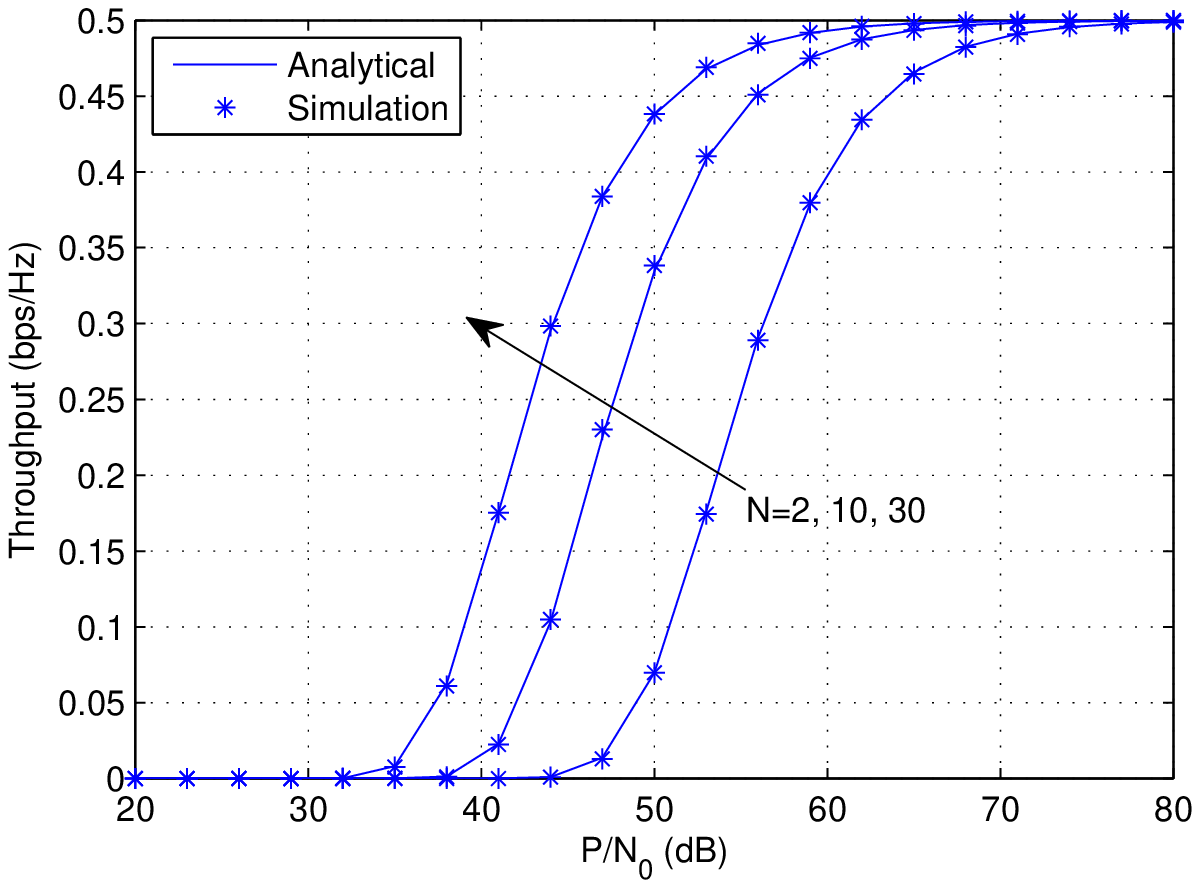}}
  \hspace{0.2in}
  \subfigure[Interference plus noise: $\tau=0.4$.]{\label{fig:2b}\includegraphics[width=0.45\textwidth]{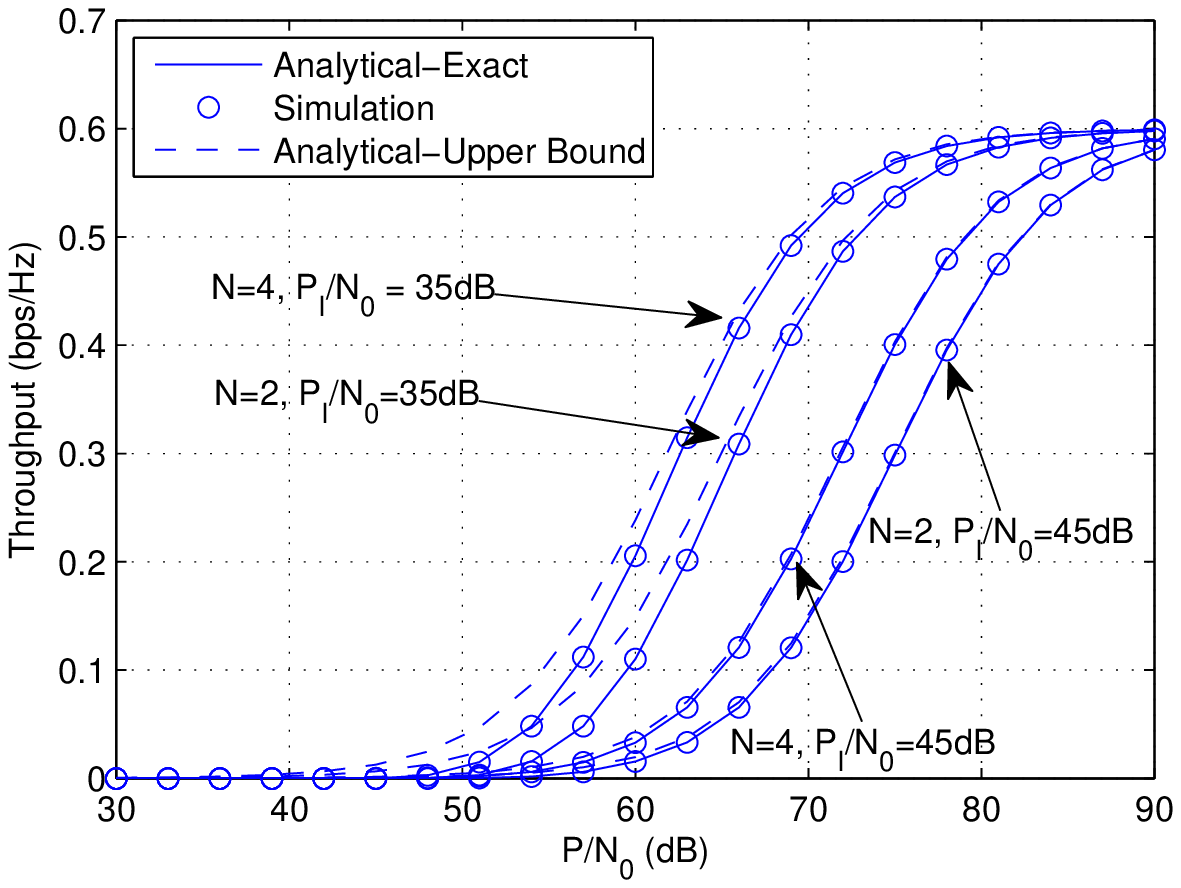}}
  \caption{Throughput of delay intolerant transmission for a fixed $\tau$ with $d_3=d_4=10\mbox{m}$.}
  \label{fig:fig2}
\end{figure}

Fig. \ref{fig:fig2} illustrates the throughput of delay intolerant transmission with a fixed $\tau$ for both the noise limited case and the interference plus noise case. In both figures, it can be readily observed that putting more antennas at the PB can significantly improve the achievable throughput. However, when the transmit power is sufficiently large, the benefit of adding extra antennas quickly diminishes. This phenomenon is quite intuitive, since increasing the number of antennas can provide higher energy beamforming gain, hence, the amount of the harvested energy at the source improves, which in turn reduces the outage probability of the system. On the other hand, with sufficiently large transmit power, the system is no longer energy limited. In this regime, the throughput performance is limited by the constant transmission rate $R_c$. For instance, given $R_c=1$, with a fixed $\tau =0.5$ or $0.4$, the maximum achievable throughput $R_c(1-\tau)$ is $0.5$ and $0.6$, respectively, as shown in the Fig. \ref{fig:fig2}. In addition, we see that the analytical upper bound (\ref{eqn:coro:5}) remains sufficiently tight, especially when the interference is strong, where the two curves overlap.

\begin{figure}[ht]
  \centering
  \subfigure[Noise only: $\tau=0.2, 0.5$.]{\label{fig:3a}\includegraphics[width=0.45\textwidth]{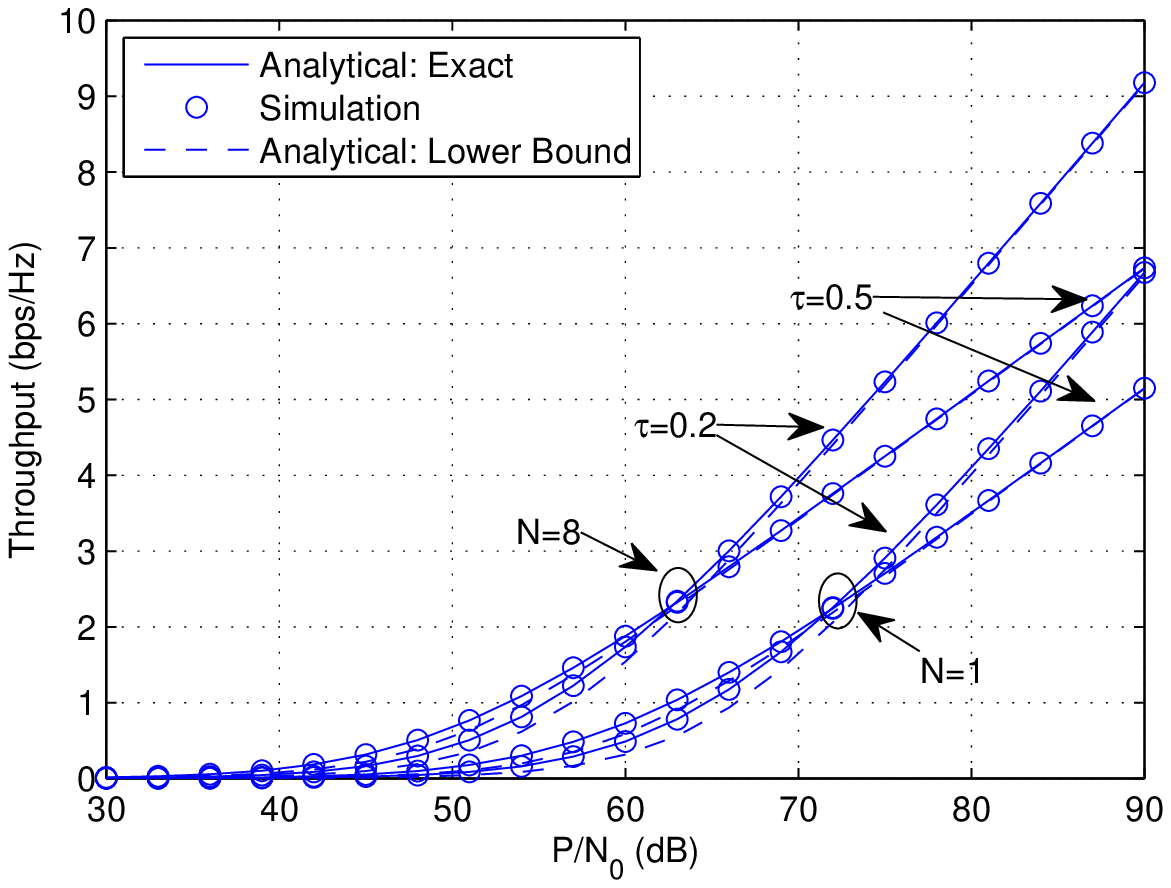}}
  \hspace{0.2in}
  \subfigure[Interference plus noise: $\tau=0.5$, $P_I/N_0=10$dB.]{\label{fig:3b}\includegraphics[width=0.45\textwidth]{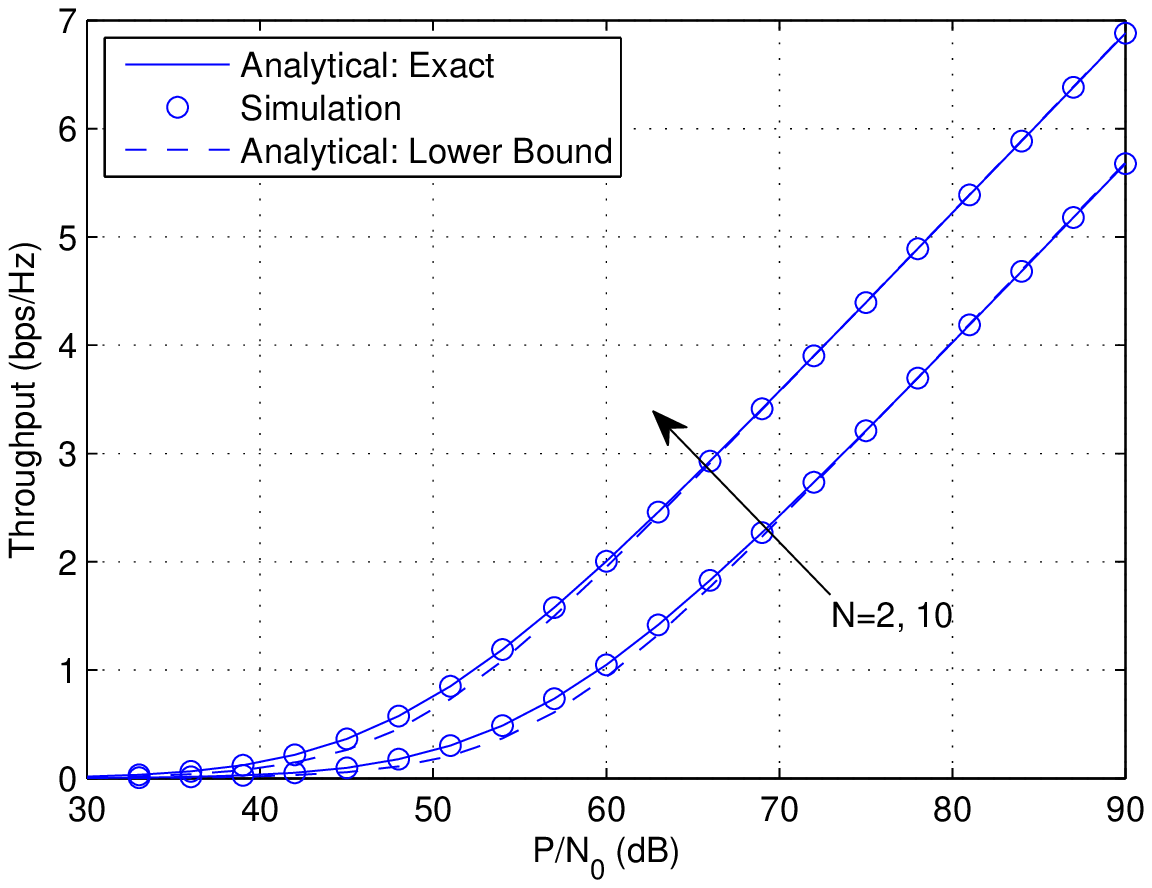}}
  \caption{Throughput of delay tolerant transmission for fixed $\tau$ with $d_3=10\mbox{m}$ and $d_4=20\mbox{m}$.}
  \label{fig:fig3}
\end{figure}

Fig. \ref{fig:fig3} shows the throughput of delay tolerant transmission with a fixed $\tau$ for both the noise limited and interference plus noise cases. Similar to the delay intolerant case, we observe that increasing the number of antennas improves the average throughput. In addition, the analytical lower bound (\ref{eqn:lower1}) and (\ref{eqn:lower2}) are quite tight and tends to the exact values when the transmit power is sufficiently large, i.e., $P/N_0\geq 70\mbox{ dB}$. Moreover, the choice of $\tau$ has a significant impact on the achievable throughput. At low transmit power level, a larger $\tau$ is preferred, while at high transmit power level, the opposite holds. This is also intuitive, since to guarantee reliable information transmission, i.e., to maintain low outage probability, a minimum amount of source energy is required. As such, when the transmit power level is low, it is desirable to spend more time to harvest energy, when the transmit power is high, only a smaller portion of time is needed for energy harvesting.

\begin{figure}[ht]
  \centering
  \subfigure[Optimal $\tau^*$ as in (\ref{eqn:largesnr}).]{\label{fig:4a}\includegraphics[width=0.45\textwidth]{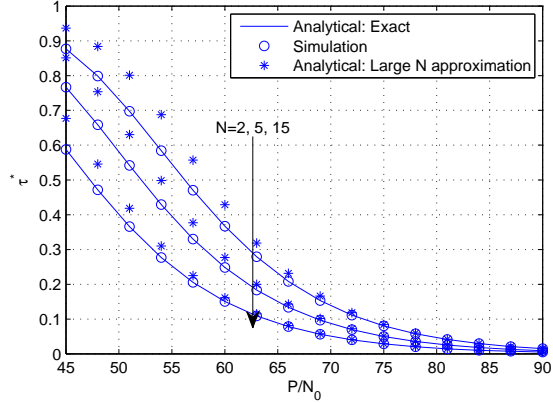}}
  \hspace{0.2in}
  \subfigure[Optimal $\tau^*$ as in (\ref{eqn:largeN}).]{\label{fig:4b}\includegraphics[width=0.45\textwidth]{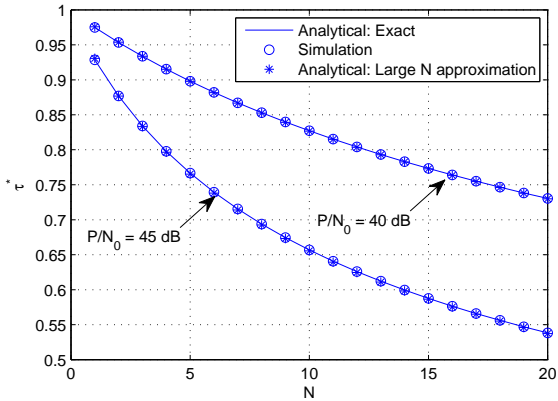}}
  \caption{Verification of the analytical expressions for the optimal $\tau^*$.}
  \label{fig:fig4}
\end{figure}

Fig. \ref{fig:fig4} examines the accuracy of the analytical approximations of the optimal $\tau^*$. As illustrated in Fig. \ref{fig:4a}, the high $P/N_0$ approximation according to (\ref{eqn:largesnr}) is quite tight for moderate transmit power level. In addition, the accuracy of the approximation improves when either the transmit power becomes large or the number of antennas increases. Fig. \ref{fig:4b} investigates the accuracy of the large $N$ approximation according to (\ref{eqn:largen}). Surprisingly, we see that, regardless of the transmit power level, the approximation performs extremely well, even for small number of antennas, i.e., $N=2$.

\begin{figure}[htb!]
\centering
\includegraphics[scale=0.6]{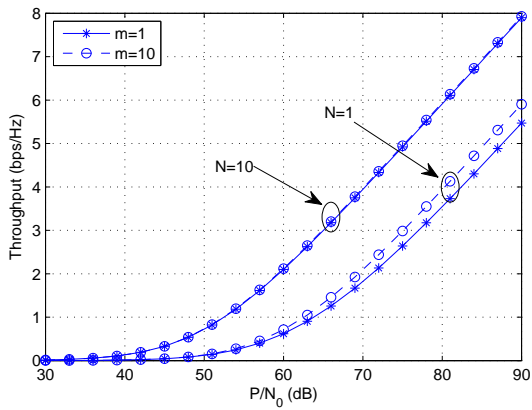}
  \caption{Impact of Nakagami-m fading parameter $m$ on the achievable throughput of delay intolerant transmission for different $N$ with $\tau=0.4$.}\label{fig:fig80}
\end{figure}

Fig. \ref{fig:fig80} investigates the impact of Nakagami-m fading parameter $m$ on the achievable throughput of delay intolerant transmission. As can be readily observed, when the number of PB antennas $N$ is small, i.e., $N=1$, a larger $m$ leads to a higher throughput. While for relatively large $N$, i.e., $N=10$, the impact of $m$ on the achievable throughput is marginal. The reason is that increasing $N$ mitigates the effect of channel fading due to the channel hardening phenomenon. As such, the benefit by increasing $m$ is substantially reduced.

\begin{figure}[htb!]
\centering
\includegraphics[scale=0.6]{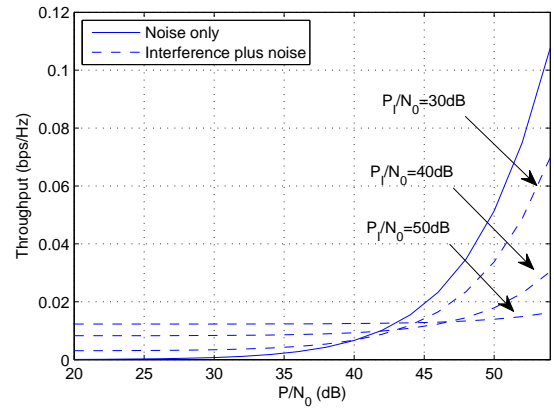}
  \caption{Achievable throughput of delay tolerant transmission with optimized $\tau$: $N=2$, $d_3=5\mbox{m}$ and $d_4=15\mbox{m}$.}\label{fig:fig5}
\end{figure}

Fig. \ref{fig:fig5} investigates the impact of interference power $P_I/N_0$ on the system throughput with optimized $\tau$. We observe an interesting phenomenon that the system throughput in the presence of strong interference is larger than that of the noise limited case when the transmit power is low. However, when the transmit power becomes large, the strong interference results in a significant throughput loss. This phenomenon is a clear demonstration of the two-sided effect of interference in wireless powered communications systems. On one hand, the interference is beneficial when exploited as an additional source of energy, and this effect is most pronounced when the transmit power is low. On the other hand, the interference is detrimental since it corrupts the information signal.


\begin{figure}[htb!]
\centering
\includegraphics[scale=0.6]{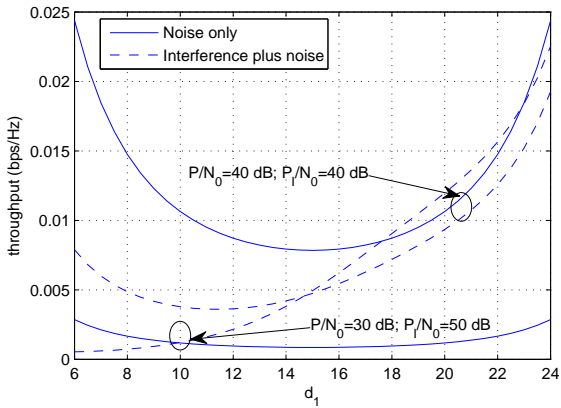}
\caption{Impact of source position on the achievable system throughput with optimized $\tau$ when $N=6$ and $m=4$.}\label{fig:fig11}
\end{figure}

Fig. \ref{fig:fig11} investigates the impact of source position on the achievable throughput of the system. In the simulations, we assume that the PB, S, and D are located on a line, and the distance from PB to D is $d_1+d_2=30 \mbox{ m}$. In the presence of interference, we assume that the distance between I and D is $d_4=15\mbox{m}$, and the included angle between sides I-D and PB-D is $\theta=\pi/6$. Hence, the distance between I and S can be computed via $d_3 = \sqrt{d_2^2+d_4^2-2d_2d_4\cos(\theta)}$. We observe that, for the noise only scenario, the throughput is a symmetric function of $d_1$ and $d_2$, and interestingly, we see that the worst case appears when S is located at the middle point. This can be explained by the fact that the effective SNR is determined by $d_1^\alpha d_2^\alpha$ as shown in (\ref{snr:noise}) which attains the maximum value when $d_1=d_2=15$. On the other hand, in the presence of interference, the throughput is no longer a symmetric function of $d_1$ and $d_2$, and the optimal location of the S depends on both the source transmit power and interference power.

\begin{figure}[htb!]
\centering
\includegraphics[scale=0.6]{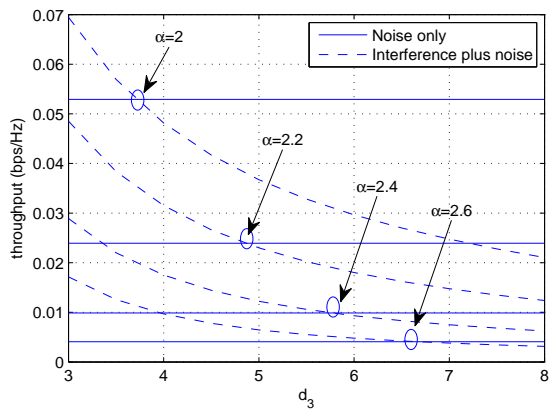}
\caption{Achievable system throughput with optimized $\tau$ for different $\alpha$ when $N=2$ and $d_3+d_4=20\mbox{m}$.}\label{fig:fig6}
\end{figure}

Fig. \ref{fig:fig6} studies the joint effect of path loss exponent $\alpha$ and interference distances on the achievable throughput for the delay tolerant transmission mode with $P_I/N_0=30 \mbox{ dB}$, $P/N_0 = 30\mbox{ dB}$ and $d_3+d_4=20\mbox{ m}$. The throughput curves associated with the noise only scenario are also plotted for comparison. We can readily observe that as $d_3$ becomes larger, the achievable throughput is reduced. This is rather intuitive since when $d_3$ increases, the energy harvested at the source from the interference signal decreases, while the interference inflicted at the destination becomes more severe. Moreover, we see an interesting behavior that, for different $\alpha$, the range of $d_3$ over which interference has a positive impact on the average throughput varies. The larger the $\alpha$, the wider the range. This can be explained as follows: For a given transmit power level, when $\alpha$ becomes larger, the amount of energy received at the source is significantly reduced, as such the extra energy contribution from the interference signal becomes more critical. In addition, with a larger $\alpha$, the detrimental effect of the interference caused at the destination is substantially diminished. Hence, even if the distance between the interferer and the source increases, the benefit of the supplying additional energy to the source still overweights the negative effect of causing the degradation of the information signal.

\begin{figure}[htb!]
\centering
\includegraphics[scale=0.6]{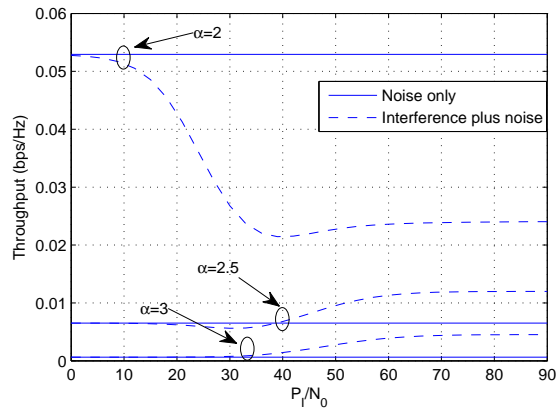}
\caption{Achievable system throughput with optimized $\tau$ for different $\alpha$ when $N=2$, $d_3 =8\mbox{m}$ and $d_4=15\mbox{m}$.}\label{fig:fig7}
\end{figure}

Fig. \ref{fig:fig7} shows the joint effect of path loss exponent $\alpha$ and interference power on the achievable throughput for the delay tolerant transmission mode with $P/N_0=40\mbox{ dB}$. As we can readily observe, whether the interference has a positive effect on the achievable throughput is closely related to the value of $\alpha$. For $\alpha=2$, the interference appears to be an undesirable factor. However, for $\alpha=2.5, 3$, the presence of strong interference could actually improve the system throughput. Moreover, a close observation of the curves associated with $\alpha=2.5$ reveals that, the effect of interference on the system throughput also depends heavily on the interference power.

\section{Conclusion}
This paper considered a point-to-point wireless powered communication system, which may find potential applications in future medical, sensor, and underwater communications systems. A detailed investigation on the average throughput of such systems was presented. For both delay intolerant and delay tolerant transmission modes, analytical expressions for the average throughput were derived, which provided efficient means for the evaluation of the average throughput. In addition, the optimal time split maximizing the average throughput was examined, and closed-form approximations were obtained, which were shown to be very accurate. Since the optimal time split does not depend on the instantaneous channel state information, it is a low complexity solution to enhance the system throughput. Finally, the impact of co-channel interference on the average throughput was studied, and the findings suggest that whether the co-channel interference will exert a positive or negative impact on the average throughput depends on the propagation environment and system setup, i.e., the path loss exponent, transmit power, network topology, and interference power.

Wireless powered communications is a newly emerged area, and there are still many theoretical and practical challenges to be tackled. For instance, the minimum required energy to activate the circuit is an important constraint to be taken into consideration; the scenario where energy storage is available at the mobile is highly relevant; the multiuser scenario \cite{Wu}, which can better exploit the available wireless power, is also an interesting topic for future study.

\appendices
\section{Proof of Lemma \ref{lemma:2}}\label{app:lemma:2}
The first derivative of $g(x)$ with respect to $x$ can be computed as
\begin{align}
g'(x) = -\exp\left(-\frac{b}{t}\right)+\frac{b(1-x)}{x^2}\exp\left(-\frac{b}{t}\right).
\end{align}
Setting $g'(x)=0$, the root of interest is
\begin{align}
x^* = \frac{\sqrt{b^2+4b}-b}{2},
\end{align}
it is easy to show that $0<x^*<1$. Now, the second derivative of $g(x)$ is
\begin{multline}
g''(x) = -\frac{b}{x^2}\exp\left(-\frac{b}{t}\right)+\frac{b^2(1-x)}{x^4}\exp\left(-\frac{b}{t}\right)\\+\left(\frac{a}{x^2}-\frac{2a}{x^3}\right)\exp\left(-\frac{b}{t}\right).
\end{multline}
Solving $g''(x)=0$ results in
\begin{align}
x^{\star} = \frac{b}{b+2}.
\end{align}
To this end, it is not difficult to show that $g''(x)<0$ when $x>x^{\star}$; and $g''(x)>0$ when $x<x^{\star}$. Hence, $g(x)$ is a convex function when $0<x\leq x^{\star}$, and $g(x)$ is a concave function when $x^{\star}<x<1$.

Observing that $g(x)\rightarrow 0$ when $x\rightarrow 0$, hence, in the regime of $0<x\leq x^{\star}$, the maximum of $g(x)$ is attained at point $x=x^{\star}$.
Now, let us turn to the second regime $x^{\star}<x<1$. To start, we find it is important to understand the relationship between $x^{\star}$ and $x^*$.
\begin{align}
\frac{x^{\star}}{x^*} &= \frac{2b}{(b+2)(\sqrt{b^2+4b}-b)}\notag\\
&= \frac{2b(\sqrt{b^2+4b}+b)}{(b+2)({b^2+4b}-b^2)}\\
&<\frac{2b+2}{2(b+2)}<1.
\end{align}
Hence, $x^{\star}<x^*$. Recall that $g(x)$ is a concave function in the second regime, the maximum of $g(x)$ is achieved at point $x=x^*$.

\section{Proof of Lemma \ref{lemma:3}}\label{app:lemma:3}
The second derivative of $g(x)$ with respect to $x$ could be computed as
\begin{align}
g''(x) =\frac{b^2}{(x-1)((b-1)x+1)^2},
\end{align}
which is strictly smaller than zero. Hence, $g(x)$ is a concave function with respect to $x$. As such, there exist an unique $x^*$ which maximizes $g(x)$, and can be obtained by solving the following equation.
\begin{align}
g'(x) =\frac{b+\frac{bx}{1-x}}{1+\frac{bx}{1-x}}-\ln\left(1+\frac{bx}{1-x}\right)=0.
\end{align}
Now, make a change of variable $y =1+ \frac{bx}{1-x}$, we have
\begin{align}
b-1+y = y\ln y,
\end{align}
After some simple algebraic manipulations, we get
\begin{align}
\frac{b-1}{e} = e^{\ln\frac{y}{e}}\ln\frac{y}{e}.
\end{align}
Recalling the definition of Lambert W function, $y$ can be expressed as
\begin{align}
y=e^{W\left(\frac{b-1}{e}\right)+1}.
\end{align}
Hence, $x^*$ could be solved as
\begin{align}
x^* = \frac{e^{W\left(\frac{b-1}{e}\right)+1}-1}{b+e^{W\left(\frac{b-1}{e}\right)+1}-1}.
\end{align}

\section{Proof of Proposition \ref{prop:3}}\label{app:prop:3}
To get the average throughput, the main task is to obtain the outage probability of the system. Hence, we start by looking at the end-to-end SINR, which can be alternatively expressed as
\begin{align}
\gamma_I =b_2UV
\end{align}
where $U=\frac{|g|^2}{|f_2|^2+b_1}$, and $V=\|{\bf h}\|^2\rho_1+
|f_1|^2\rho_I$. To characterize the statistical distribution of $\gamma_I$, the distributions of random variables $U$ and $V$ are required. We start with $U$. The cdf of $U$ can be written as
\begin{align}
F_U(x) & = {\tt Pr}\left\{\frac{|g|^2}{|f_2|^2+b_1}\leq x\right\}.
\end{align}
Since $|g|^2$ is an exponentially distributed random variable, conditioned on $|f_2|^2$, we have
\begin{align}
F_U(x||f_2|^2)&=1-e^{-b_1x}e^{-|f_2|^2x}.
\end{align}
Averaging over $|f_2|^2$, the unconditional cdf can be obtained as
\begin{align}
F_U(x)&=1-e^{-b_1x}\int_0^{\infty}e^{-(1+x)y}dy\\
&=1-\frac{e^{-b_1x}}{1+x}
\end{align}

Now, taking into account that $V$ is the sum of two independent random variables following chi-square distribution and exponential distribution, respectively, the characteristic function of $V$ can be computed as
\begin{align}
\Phi_V(s) = \frac{1}{(\rho_1s/m+1)^{Nm}(\rho_Is+1)}.
\end{align}
To this end, taking the inverse Laplace transform with the help of \cite[Eq. 5.2.21]{Integral} yields,
\begin{align}\label{pdf:v}
f_v(x)& = \frac{m^{Nm}}{\rho_1^{Nm}\rho_I}\left(\sum_{t=1}^{Nm}\frac{(-1)^{t-1}}{(Nm-t)!}\left(\frac{1}{\rho_I}-\frac{m}{\rho_1}\right)^{-t}x^{Nm-t}e^{-\frac{mx}{\rho_1}}\right.\notag\\
&\left.+\left(\frac{m}{\rho_1}-\frac{1}{\rho_I}\right)^{-Nm}e^{-\frac{x}{\rho_I}}\right).
\end{align}

Hence, the exact outage probability can be computed as
\begin{align}
&P_{\sf out}={\tt E}\left\{F_U\left(\frac{\gamma_{\sf th}}{b_2V}\right)\right\}.
\end{align}
To this end, the desired result can be obtained after some algebraic manipulations.

\section{Proof of Proposition \ref{prop:6}}\label{app:prop:6}
Using Jensen's inequality, the ergodic capacity can be lower bounded by
\begin{align}
C_e &\geq \log_2\left(1+\exp\left\{{\tt E}\left\{\ln\gamma_I\right\}\right\}\right).
\end{align}
To evaluate ${\tt E}\left\{\ln\gamma_I\right\}$, the key is to compute the expectation of $\ln U$ and $\ln V$, which we do in the following.
\begin{align}
{\tt E}\left\{\ln U\right\} &={\tt E}\left\{\ln |g|^2\right\} -{\tt E}\left\{\ln( |f|^2+b_1)\right\} \\
&=\psi(1)+e^{b_1}{\tt Ei}(-b_1),
\end{align}
where we have used the integration formula \cite[Eq. (4.337.1)]{Table}
\begin{align}
\int_0^{\infty}e^{-\mu x}\ln(x +\beta)dx = \frac{\ln \beta -e^{\mu\beta}{\tt Ei}(-\mu\beta)}{\mu}.
\end{align}

Similarly, using the pdf of $V$ given in Eq. (\ref{pdf:v}), ${\tt E}\left\{\ln V\right\}$ can be computed as (\ref{eqn:lnv}),
\begin{figure*}[t]
\begin{multline}\label{eqn:lnv}
{\tt E}\left\{\ln V\right\}=\frac{m^{Nm}}{\rho_1^{Nm}\rho_I}\left(\sum_{t=1}^{Nm}\frac{(-1)^{t-1}}{(Nm-t)!}\left(\frac{1}{\rho_I}-\frac{m}{\rho_1}\right)^{-t}\left(\left(\frac{\rho_1}{m}\right)^{mN-t+1}\Gamma(mN-t+1)\left(\psi(mN-t+1)+\ln\frac{\rho_1}{m}\right)\right)\right.
\\\left.+\left(\frac{m}{\rho_1}-\frac{1}{\rho_I}\right)^{-Nm}\left(\rho_I\left(\psi(1)+\ln\rho_I\right)\right)\right).
\end{multline}
\hrule
\end{figure*}
where we have used the integration formula \cite[Eq. (4.352.1)]{Table}
\begin{align}
\int_0^{\infty}x^{v-1}e^{-\mu x}\ln xdx = \frac{1}{\mu^v}\Gamma(v)\left(\psi(v)-\ln \mu\right).
\end{align}
To this end, pulling everything together yields the desired result.

\nocite{*}
\bibliographystyle{IEEE}
\begin{footnotesize}

\end{footnotesize}

\end{document}